\documentclass[prb,twocolumn,showpacs,preprintnumbers,amsmath,amssymb,floats]{revtex4-1}
\usepackage{graphicx}
\usepackage{amsmath}
\usepackage{epstopdf}
\usepackage{amsbsy}
\usepackage{color}
\usepackage[version=3]{mhchem}
\usepackage{dcolumn}
\usepackage{bm}
\usepackage{amssymb}

\begin{document}

\title{Competing superconducting and magnetic order parameters and field-induced magnetism in electron doped \ce{Ba(Fe_{1-x}Co_{x})2As2}}

\author{J. Larsen$^1$, B. Mencia Uranga$^2$, G. Stieber$^2$, S. L. Holm$^{2,3}$, C. Bernhard$^{4}$, T. Wolf$^{5}$, \\ K. Lefmann$^2$, B. M. Andersen$^2$, C. Niedermayer$^3$}

\affiliation{$^1$Department of Physics, Technical University of Denmark, 2800 Kgs. Lyngby, Denmark\\
$^2$Niels Bohr Institute, University of Copenhagen, 2100 Copenhagen {\O}, Denmark\\
$^3$Laboratory for Neutron Scattering and Imaging, Paul Scherrer Institute, CH-5232 Villigen, Switzerland\\
$^4$University of Fribourg, Department of Physics and Fribourg Centre for Nanomaterials, Chemin du Mus\'{e}e 3, CH-1700 Fribourg, Switzerland\\
$^5$Karlsruher Institut f\"ur Technologie, Institut f\"ur Festk\"{o}rperphysik, D-76021 Karlshuhe, Germany}

\date{\today}

\begin{abstract}

We have studied the magnetic and superconducting properties of \ce{Ba(Fe_{0.95}Co_{0.05})2As2} as a function of temperature and external magnetic field using neutron scattering and muon spin rotation. Below the superconducting transition temperature the magnetic and superconducting order parameters coexist and compete. A magnetic field can significantly enhance the magnetic scattering in the superconducting state, roughly doubling the Bragg intensity at 13.5~T. We perform a microscopic modelling of the data by use of a five-band Hamiltonian relevant to iron pnictides. In the superconducting state, vortices can slow down and freeze spin fluctuations locally. When such regions couple they result in a long-range ordered antiferromagnetic phase producing the enhanced magnetic elastic scattering in agreement with experiments.

\end{abstract}


\maketitle

\section{Introduction}
Unconventional superconductors including cuprates, iron-based systems, and heavy fermion materials constitute interesting classes of materials that currently evade quantitative theoretical modelling\cite{scalapino12}. A fundamental outstanding question is, of course, the origin of the superconducting phase itself. For most of these materials, the superconductivity takes place in close proximity to a magnetically ordered phase which may coexist with superconductivity in a limited doping range. For the cuprates, this coexistence region has been extensively studied  experimentally and theoretically on both the electron- and hole-doped sides of the phase diagram. In the latter case, the so-called stripe order and its effects on superconductivity remain an active area of research.\cite{cupratereview1,cupratereview2}

For the iron-based materials, a magnetic $(\pi,0)$ stripe ordered phase generally precedes the superconducting phase. The emergence of a coexistence region of static magnetic and superconducting order at intermediate doping levels depends on the particular material family; whereas the 122 systems exhibit a coexistence region at the foot of the magnetic dome\cite{LumsdenJPhys2010}, this is not the case for most of the 1111 compounds (Sm-1111 being an exception.\cite{sanna09,DrewNatureMat2009}) For both hole- and electron-doped Ba-122, the coexistence region has been extensively studied and constitutes a topic of controversy with some measurements suggesting meso-scale phase separation while others indicate microscopic coexistence.\cite{aczel08,fukazawa09,park09,goko09,julien09,bernhard09,laplace09,marsik10} Initial measurements suggested that K substitution (hole doping, out-of-plane ions) led to phase separation\cite{aczel08,fukazawa09,park09,goko09} whereas more recent experiments are in agreement with 
microscopic coexistence.\cite{HChen09,wiesenmayer11,li12} Co substitution (electron doping, in-plane ions) in the underdoped regime of Ba-122 seems to produce microscopic phase coexistence where superconducting and magnetic orders compete for the same electronic states.\cite{laplace09,julien09,bernhard09,marsik10} The latter is in agreement with neutron diffraction measurements finding a reduction of the magnetic Bragg peak intensity upon entering the superconducting state,\cite{pratt09,christianson09} even though changes of the volume fraction can produce the same effect. Most recent experimental studies find that Co doped Ba-122 samples exhibit volume-full coexistence at $x \lesssim 0.04-0.06$, while above this doping level superconductivity and magnetism do not coexist microscopically, but rather consist of inhomogeneous distributions of frozen (or slowly fluctuating) magnetic islands in the superconducting host.\cite{bernhard09,bernhard12} A recent $^{75}$As nuclear magnetic resonance (NMR) study, 
consistent with this latter scenario, found an inhomogeneous cluster spin glass coexisting with superconductivity up to $x\sim 0.07$\cite{dioguardi13} in agreement with earlier muon spin rotation ($\mu$SR) measurements.\cite{bernhard09,bernhard12} A similar short-range cluster spin glass  phase has been also recently found near optimal (electron) doping in Ni-doped Ba-122.\cite{Lu14}
 
The experimental results discussed in the above section exemplifies well the fascinating parameter dependence of the iron pnictides; the properties of the resulting many-particle ground state depend on the particular compound and the specific doping method. Other recent examples of this behavior includes several Mn-substituted samples revealing unusual competitive effects between magnetism and superconductivity as well as the emergence of new high-temperature impurity-induced magnetic phases.\cite{tucker12,inosov13,MNG14,hammerath14}
 
It is important to further study the coexistence region of the iron pnictides because it can provide crucial information about the ground state properties of these systems. This is evident, for example, in the fact that the detailed superconducting gap structure inside and outside the coexistence region are closely linked.\cite{parker09,maiti12} Furthermore, theoretical studies find that the mere presence of the coexistence region is evidence for $s \pm$ pairing symmetry of the superconducting phase and incommensurability of the magnetic order.\cite{vorontsov09,vorontsov10,fernandes10} Hence, the coexistence region can be used as a guide for theoretical models, since the competition between the superconducting and magnetic orders can only be obtained with the correct gap structure, and a complete quantitative description requires correct orbital content of the order parameters as well. For the case of Ba-122 there is an additional interesting question of what happens to the two distinct transitions, 
magnetic versus nematic-structural, in the coexistence phase.\cite{nandi10,fernandes13} Theoretically, a quantitative theory of superconducting pairing from spin fluctuations in the coexistence phase is highly complex, especially for multi-band iron-based materials, due to the folding of the Fermi surfaces by the magnetic ordering. Only a limited number of microscopic studies exist for the calculation of pairing potentials within the symmetry-broken SDW phase.\cite{schrieffer89,schmidt14}

Exploring the consequences of external magnetic fields in the coexistence region constitutes a probe of the interplay between superconductivity and magnetism. In the cuprates, neutron scattering has revealed that application of an external magnetic field perpendicular to the CuO$_2$ planes causes a pronounced enhancement of the elastic incommensurate (stripe) magnetic order at doping levels in a range: $0.10 \leq x \leq 0.135.$\cite{Lake02,Chang08,Kofu09} At larger doping values, the static stripe order is absent, but can be induced by application of a magnetic field.\cite{Chang09,Khaykovich} For the iron pnictides, the influence of a magnetic field on the neutron scattering cross section is less explored at present. Previous neutron studies on Ni-doped Ba-122 mainly focussed on the effects of an external magnetic field in reducing the neutron resonance in the inelastic response.\cite{zhao10,li11,wang11} Similar results were obtained for the resonance in the superconducting state of FeSe$_{1-x}$Te$_x$.\cite{
li11,wen10} For underdoped Ba(Fe$_{0.96}$Ni$_{0.04})_2$As$_2$, the influence of an in-plane magnetic field on the static (but short-range) antiferromagnetism was investigated recently by Wang {\it et al.},\cite{wang11} who found field-enhanced static antiferromagnetic scattering below the superconducting critical transition temperature $T_c=17$ K at the expense of spectral weight at the resonance. Open questions remain, however, about the field dependence of the magnetic and superconducting volume fractions discernible only by local probes.

Here we perform a systematic neutron and $\mu$SR experimental study combined with theoretical modelling of the interplay between magnetic and superconducting order in the coexistence region of \ce{Ba(Fe_{0.95}Co_{0.05})2As2}. We verify earlier experimental studies of this composition in finding a nano-scale coexistence of magnetic and superconducting orders exhibiting strong competition as a function of temperature.\cite{bernhard12} More specifically, the magnetic moments are strongly suppressed below $T_c$. However, as a function of external magnetic field the magnetism can be recovered similar to what has been observed for the cuprates.\cite{Lake02,Chang08,Kofu09} Using a microscopic model with realistic band structure and including superconductivity and magnetic correlations, we explain the experimental observations. In particular, we show how vortices in the superconducting condensate can locally pin magnetic $(\pi,0)$ ordered regions giving rise to the field-enhanced static 
magnetic scattering in the superconducting state.

\section{Experiments}\label{ExpIntro}
At room temperature, the antiferromagnetic parent compound \ce{BaFe2As2} exhibits the so-called \ce{ThCr2Si2} crystal structure (space group 71, I4/mmm) with $a = b = 3.96$~\AA\ and $c = 13.02$~\AA.\cite{RotterPRB} This structure consists of FeAs layers which are stacked along the $c$-axis and separated by interstitial Ba ions. The material can be made superconducting by substitution of a small fraction, $x$, of the Fe ions by Co. Superconductivity emerges above x $\approx$ 0.04.

Concomitant to the magnetic transition temperature, $T_N$, the structure changes from the tetragonal to a low temperature orthorhombic (space group 69, Fmmm) phase with $\widetilde{a} = 5.61$~\AA$\:$, $\widetilde{b} = 5.57$~\AA$\:$~ and $\widetilde{c} \approx c$.\cite{HuangPRL,RotterPRB} $T_c$ and the structural transition temperature, $T_s$, decrease in very similar manners when $x$ increases. In the superconducting phase the magnetic moment is reduced to about $\thicksim 0.1 \mu_B$ per Fe$^{2+}$,\cite{bernhard12} which is about $10~\%$ of the magnetic moment of the parent compound having $\mu = 0.9-1.0~\mu_B$ per Fe,\cite{HuangPRL,LumsdenJPhys2010} which could indicate an incommensurate SDW type order. However, no evidence of incommensurability was found in this sample, the antiferromagnetic Bragg peaks are resolution limited.\cite{bernhard12,christianson09} The doping range between $0.04 \leq x \leq 0.06$ in Ba-122 is generally referred to as the coexistence region. 
Here, both neutron scattering\cite{christianson09,pratt09} and transverse field $\mu$SR\cite{bernhard12,marsik10} measurements provide clear evidence for a direct competition between the superconducting and magnetic order parameters on a nanometer scale. The magnetic order parameter is anomalously suppressed below $T_c$ and the excellent agreement between neutron diffraction and transverse field $\mu$SR, which unlike neutron diffraction is a local probe, suggests that both techniques probe the same magnetic state,\cite{bernhard12} which must then persist throughout the whole crystal. In addition, neutron diffraction excludes the possibility that the suppression of the magnetic order parameter could be due to a magnetic phase transition or a spin reorientation.\cite{christianson09}

Infrared optical spectroscopy performed on \ce{Ba(Fe_{0.945}Co_{0.055})2As2} clearly shows that the superconducting energy gap exists for all electronic states\cite{marsik10} at low temperature. At the same time, the magnetic volume fraction extracted from $\mu$SR measurements are found to be $\thicksim 80-90~\%$,\cite{marsik10,bernhard12} see also Fig.~\ref{MuonData} in this paper. If superconductivity and magnetism had phase separated into well-defined macroscopic regions of the sample, the non-superconducting regions would have appeared in the optical spectroscopy. Similarly, if the suppression of the magnetic signal observed by neutron diffraction had been due to a change in magnetic volume fraction, we would expect the volume fraction to be lower than $\thicksim 80-90~\%$. In addition, the temperature dependence of the so-called pair breaking peak, observed in optical spectroscopy, provides further evidence for the competition between superconductivity 
and magnetism.

In this section, we present experimental work done on a $\thicksim 1$~g  single crystal of \ce{Ba(Fe_{0.95}Co_{0.05})2As2} produced by our laboratory in Karlsruhe. Resistivity measurements (not shown here) shows that $T_c \approx 20$~K and previously published $\mu$SR measurements (on the same sample) shows $T_N \approx 40 $~K,\cite{bernhard12} consistent with earlier published values for this material.\cite{NiPRB2008,ChuPRB2009}

\subsection{Neutron Scattering Experiment}
The neutron scattering data presented in this paper were measured at the cold neutron triple axis spectro\-meter RITA-II \cite{Lefmann06,Bahl06} located at the Paul Scherrer Institut (PSI), Switzerland. The spectrometer was operated under elastic conditions ($\hbar\omega = 0$) in a monochromatic $\mathbf{Q}$ dispersive mode with $k_i=k_f = 1.49$~\AA$^{-1}$ giving an energy resolution of $0.2$~meV (FWHM). To clean the signal of higher order contamination a PG-filter was placed before the sample and a Be-filter was placed in the outgoing beam. An 80’ collimator was placed after the monochromator and the secondary spectrometer has a natural collimation of 40’ for this sample size. The scattered neutrons were detected using a 128x128 pixels PSD\cite{Bahl06}.

We measured both longitudinal and transverse scans in reciprocal space around the magnetic Bragg peak at $(\frac{1}{2}, \frac{1}{2}, 3)$ (all $\mathbf{Q}$-positions in the experimental section are in tetragonal notation). To obtain these scan directions, the sample was mounted in two different orientations; the longitudinal orientation, having the $(1, 1, 0)$ and $(0, 0, 2)$ peaks in the scattering plane and the transverse orientation, having the $(1, 1, 6)$ and $(\bar{1}, 1, 0)$ peaks in the scattering plane. To be able to apply high magnetic fields and access the temperature range necessary for this study, the sample was mounted in a $15$~T Oxford Instruments vertical field cryo-magnet. For the longitudinal scans the magnetic field was thus applied along $(\bar{h}, h, 0)$. For the transverse scans the field was along the $(3h, 3h, \bar{h})$ direction.

\begin{figure}[t]
\centering
	\includegraphics[width=0.45\textwidth]{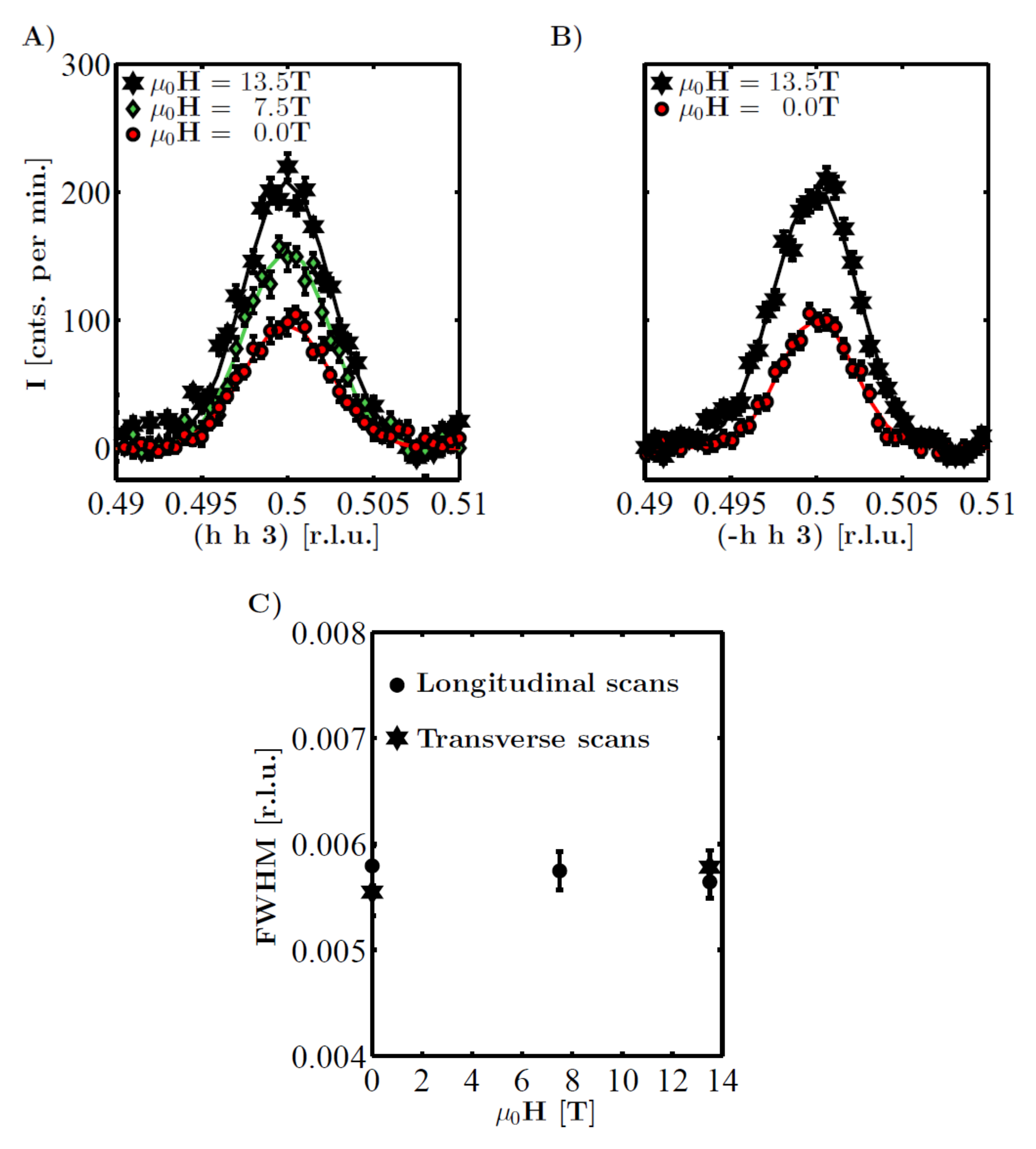}
	\caption{(Color online) (A) Elastic $\mathbf{Q}$-scans of the magnetic $\mathbf{Q}=(\frac{1}{2}, \frac{1}{2}, 3)$ Bragg peak, performed in the longitudinal orientation i.e. along the $(h, h, 3)$ direction in reciprocal space which is equivalent to $(h, 0 , 3)$ in the orthorhombic notation. The figure shows scans at $\mu_{0}H_{} = 0$~T, $7.5$~T and $13.5$~T. The solid lines represents fits to a single Gaussian peak. (B) Elastic $\mathbf{Q}$-scans of the magnetic $\mathbf{Q}=(\frac{1}{2}, \frac{1}{2}, 3)$ Bragg peak, made in the transverse orientation i.e. along the $(h, \bar{h}, 3)$ direction in reciprocal space, which is equivalent to $(0, k, 3)$ in the orthorhombic notation. The figure shows scans made at $\mu_{0}H_{} = 0$~T and $13.5$~T. (C) The Gaussian FWHM obtained in both orientations as function of field.}
	\label{CompareBraggPeaks}
\end{figure}

Fig.~\ref{CompareBraggPeaks} A shows the magnetic Bragg peak at $\mathbf{Q} = (\frac{1}{2}, \frac{1}{2}, 3)$ at 0~T, 7.5~T and 13.5~T scanned in the longitudinal orientations at 2~K, and Fig.~\ref{CompareBraggPeaks} B shows the magnetic Bragg peak at $\mathbf{Q} = (\frac{1}{2}, \frac{1}{2}, 3)$ at 0~T and 13.5~T scanned in the transverse orientation at 2~K. We find that the line shape is essentially unaltered by the applied field for both orientations, i.e. neither the peak position nor the peak width changes significantly, see Fig.~\ref{CompareBraggPeaks} C. These data suggests that the only field dependent parameter is the peak intensity, which is seen to increase significantly as the field increases.

\begin{figure}[t]
\centering
	\includegraphics[width=0.45\textwidth]{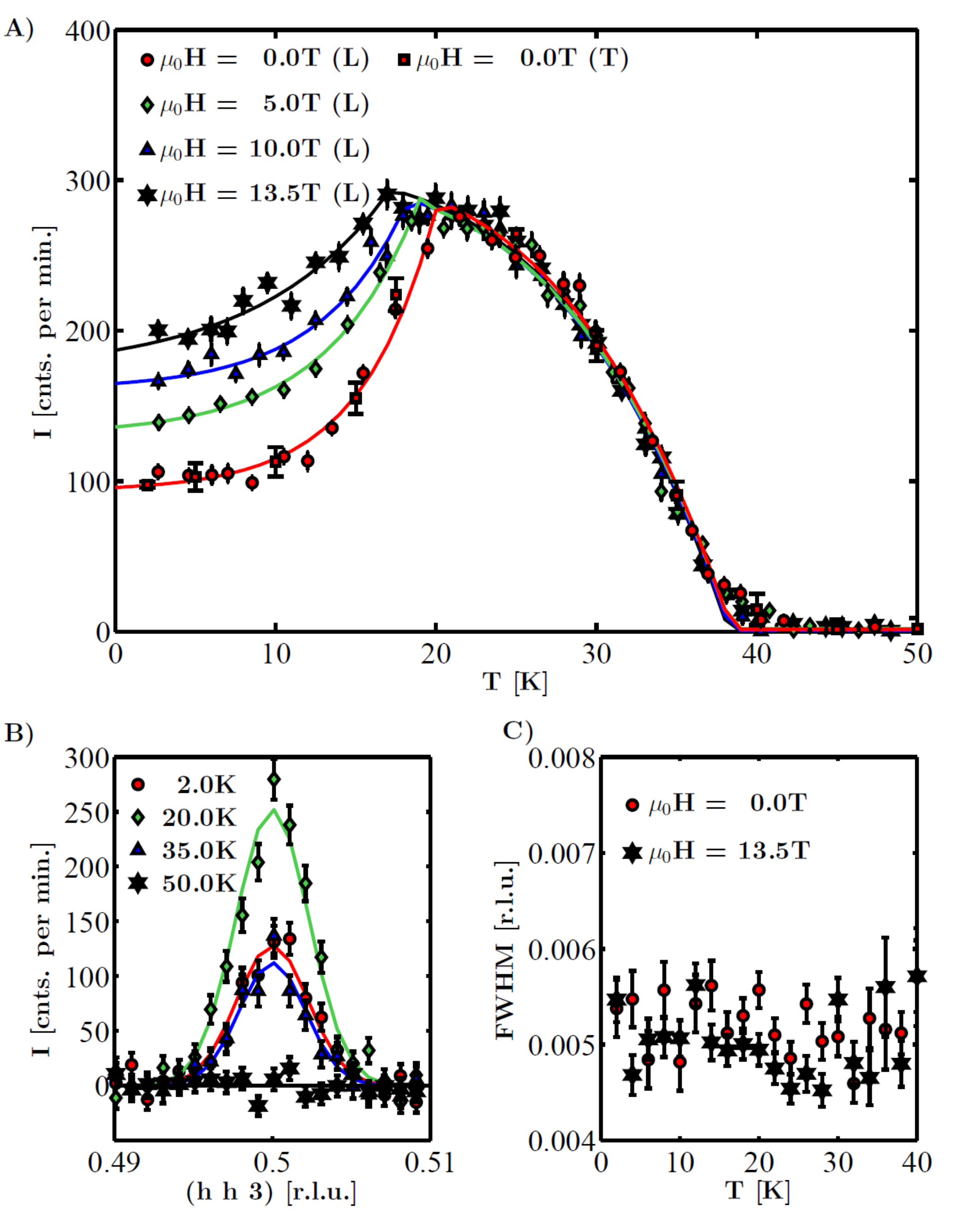}
	\caption{(Color online) (A) Intensity of the $\mathbf{Q}=(\frac{1}{2}, \frac{1}{2}, 3)$ Bragg peak measured by elastic neutron scattering in the longitudinal orientation (L) and the transverse orientation (T). The SDW intensity is seen to increase at low temperature upon applying a magnetic field. The solid lines are guides to the eyes. (B) Elastic $\mathbf{Q}$-scans of the $\mathbf{Q}=(\frac{1}{2}, \frac{1}{2}, 3)$ Bragg peak at temperatures 2~K, 20~K, 35~K and 50~K in zero field. No changes in the peak position is observed. The solid lines are fits of a single Gaussian peak on a flat background. The background is determined from the 50~K scan. (C) The Gaussian FWHM as function of temperature below $T_N$ at $\mu_{0}H_{} = 0$~T and $13.5$~T.}
	\label{CompetitionOrder}
\end{figure}

In Fig.~\ref{CompetitionOrder} we present the temperature dependence of the magnetic Bragg peak intensity at $\mathbf{Q}=$($\frac{1}{2}, \frac{1}{2}, 3$) in zero field. Above $T_{N}$, the intensity is zero, and cooling below $T_{N}$, the intensity gradually develops, typical for a second order magnetic phase transition. The magnetic signal has the largest intensity right at $T_{c}$ and becomes increasingly suppressed, upon entering the superconducting phase, reaching a plateau-like value at low temperatures, see Fig.~\ref{CompetitionOrder} A and B. In addition to the data at zero field, Fig.~\ref{CompetitionOrder} A also shows data measured in 5.0~T, 10.0~T and 13.5~T. Data above $T_c$ are remarkably similar for all the field values, indicating that the magnetic phase itself is unaltered by the field. However, the degree to which the magnetic intensity is suppressed in zero field is significantly reduced for all higher field values, corresponding to a recovery of some of the magnetic order parameter otherwise 
lost inside the superconducting phase. Independent of the scan orientation, at the highest field probed here, around half of the lost low-temperature signal is regained.

Fig.~\ref{CompetitionOrder} C shows the temperature dependence of the Gaussian peak width in zero field and in 13.5~T respectively. We observe no significant changes of the peak width, indicating the effect which suppresses the magnetic peak intensity in high fields is the same as in zero field, however, less pronounced. Our measurements in zero field are in agreement with the findings in Christianson {\it et al.},\cite{christianson09} who also found no significant effect of the appearance of the superconducting state on the SDW Bragg peak.

\subsection{Muon Spin Rotation Experiment}
The $\mu$SR measurements presented in this paper were performed with the general purpose spectrometer (GPS) at the Swiss Muon Source (S$\mu$S) at PSI. The $\mu$SR technique measures the time-resolved spin depolarization of an ensemble of muons inside a material. A beam of spin polarized muons is implanted in the sample where they come to rest without any significant loss in spin polarization. Among the pnictide materials the muons are known to stop at well-defined interstitial lattice sites close to an As ion\cite{MaeterPRB09}. The average muon implantation depth is about 100 $\mu$m. The magnetic and superconducting properties probed by the muon ensemble are therefore representative of the bulk sample.

The mean life-time of positive muons is $\tau \approx 2.2~\mu$s, after which it decays, $\mu^{+} \rightarrow e^{+} + \nu_{e} + \bar{\nu}_{\mu}$, where the positron is preferentially emitted along the direction of the muon spin at the instant of the decay. By recording the time evolution of the spatial asymmetry of the positron emission rate, $P(t)$, the time-resolved spin polarization of the muon ensemble is obtained. Before the decay, the muon precesses in the local field, $B_{\textrm{loc}} = \mu_{0}H_{\textrm{loc}}$, arising from the sum of the applied field and the dipole field of the nearby magnetic ions. The spin precession frequency is given by $\omega_{\mu} = \gamma_{\mu} B_{\textrm{loc}}$ where $\gamma_{\mu} = 851.4$~MHz/T is the gyromagnetic ratio of positive muons.

\begin{figure}[b]
	\centering{\includegraphics[width=0.48\textwidth]{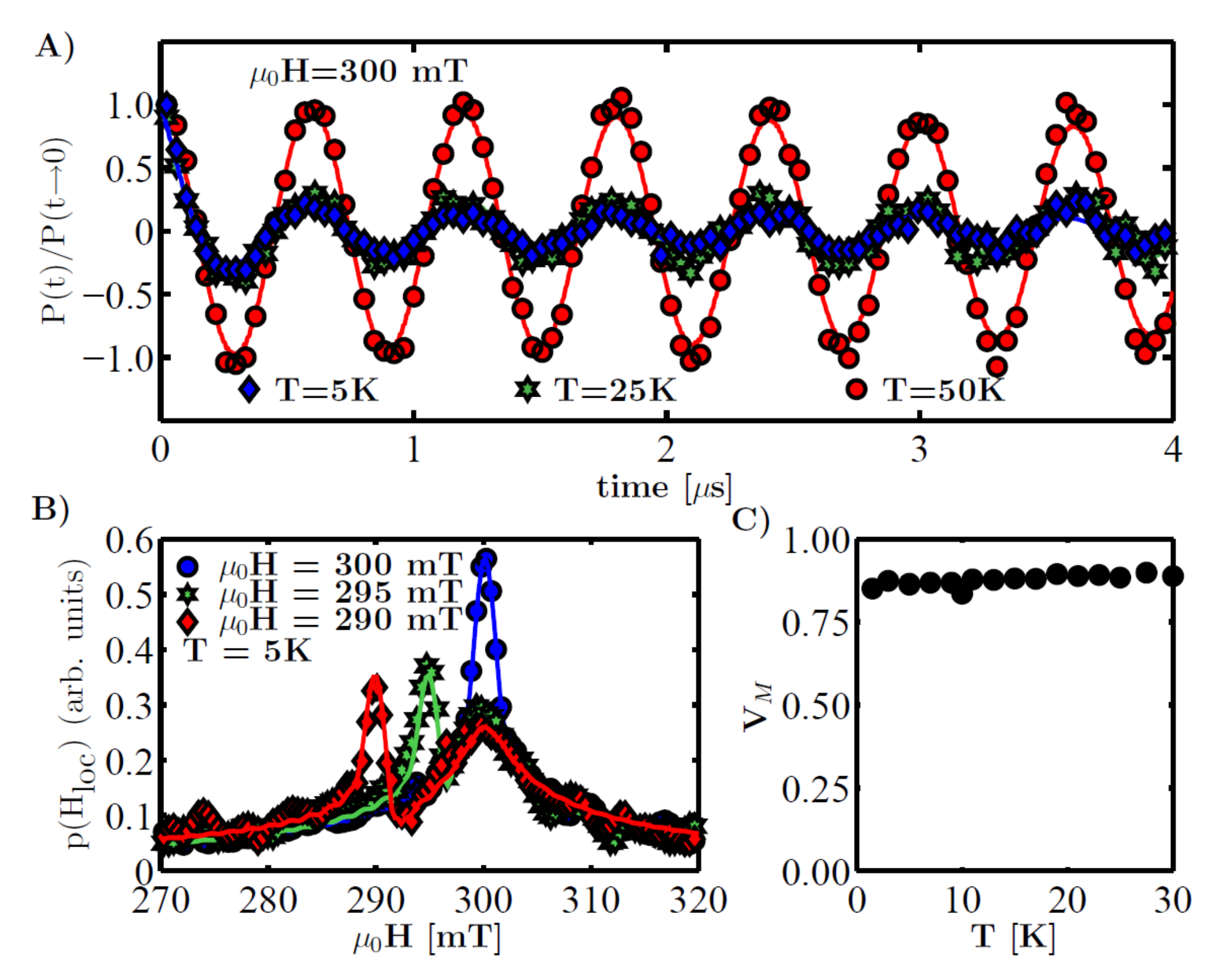}}
	\caption{(Color online) (A) Transverse field $\mu$SR time spectra. Below $T_c$ we observe a magnetic component from the sample and a non-magnetic component which we believe to be from muons stopping outside the sample. (B) $\mu$SR line shapes from the pinning experiment, showing one narrow peak corresponding to the external field in the otherwise non-magnetic region, and one broad peak corresponding to magnetic flux trapped as pinned vortices coexisting with antiferromagnetism. (C) Temperature dependence of the magnetic volume fraction at $\mu_{0}H_{} = 300$~mT below 30~K.}
	\label{MuonData}
\end{figure}

Fig.~\ref{MuonData} A shows the outcome of our transverse field $\mu$SR measurement. These spectra were recorded with an external field of $\mu_{0}H_{} = 300$~mT applied along the $c$-axis of the sample and transverse to the initial spin direction of the muon beam. We obtain excellent fits of the data below $T_c$ using the two component model Eq. (\ref{eq:muonpol}). The first component, $A_{f}$, is rapidly relaxing and evidently originates from the vast majority of muons stopping in a region of the sample with magnetic order, and the small second component, $A_s$, which is essentially not relaxing and originates from muons stopping in a non-magnetic region, probably outside the sample.\cite{bernhard12} The two-component model reads
\begin{eqnarray} \label{eq:muonpol} \nonumber
	P(t) &=& P(0) \left[ A_{f} \cos ( \omega_{f} t + \phi ) \exp( - \lambda_{f} t ) \right. \\
	     &\:& \phantom{P(0)}  + \left. A_{s} \cos ( \omega_{s} t + \phi ) \exp( - \lambda_{s} t ) \right],
\end{eqnarray}
where $P(0)$ is the initial asymmetry and the muon spin precession frequencies $\omega_{f}$ and $\omega_{s}$ are related to the local field in magnetic and non-magnetic regions respectively. $\phi$ is the initial phase of the muon spin. A similar model was used in previous analyses.\cite{marsik10,bernhard12} The temperature dependence of the transverse field relaxation rate, $\lambda_f$, published by Bernhard {\it et al.},\cite{bernhard12} clearly demonstrates the same suppressed magnetic signal as seen by neutron diffraction. We find a magnetic volume fraction appearing below $T_N = 40$ K, which occupies around $90\%$ (effectively probably 100$\%$) of the sample even below $T_c$ all the way down to the lowest temperature probed, see Fig.~\ref{MuonData} C. These results are in good agreement with previous findings.\cite{marsik10,bernhard12}

Fig.~\ref{MuonData} B shows the result of a so-called pinning experiment\cite{marsik10} displaying $\mu$SR-line shapes, as obtained from Fourier transformation of the transverse field $\mu$SR time spectra, showing the distribution of magnetic fields probed by the muon ensemble, $\textrm{p}(\textrm{H}_{\textrm{loc}}$). The measurements are performed after field cooling the sample in 300~mT and subsequently then reducing the field to 295~mT and 290~mT, respectively. The line shapes consist of a narrow peak which represents muons stopping in the non-magnetic environment which only probes the external field distribution, and a very broad peak due muons stopping inside a magnetic environment which probe the pinned magnetic distribution from the coexisting SDW and superconductor state. The broad peak remains fully pinned when $\mu_{0}H_{}$ is changed. This is evidence of complete pinning of the trapped magnetic flux vortices in the sample, an effective pinning effect like the one displayed by this sample, is a 
hallmark of a bulk type-II superconductor with a strongly pinned vortex lattice.\cite{marsik10,bernhard12}

\section{Theoretical Description}

\subsection{Model}
The starting point for the theoretical modelling is the following five-orbital Hamiltonian
\begin{equation}
 \label{eq:H}
 H=H_{0}+H_{int}+H_{BCS}+H_{Z},
\end{equation}
where $H_0$ contains the kinetic energy given by a tight-binding fit to the DFT bandstructure\cite{ikeda10} including hopping integrals of all Fe orbitals to fifth nearest neighbors
\begin{equation}
 \label{eq:H0}
H_{0}=\sum_{\mathbf{ij},\mu\nu,\sigma}t_{\mathbf{ij}}^{\mu\nu}e^{i \varphi_{\mathbf{ij}} }c_{\mathbf{i}\mu\sigma}^{\dagger}c_{\mathbf{j}\nu\sigma}-\mu_0\sum_{\mathbf{i}\mu\sigma}n_{\mathbf{i}\mu.\sigma}.
\end{equation}
Here, the operators $c_{\mathbf{i} \mu\sigma}^{\dagger}$ create electrons at the $i$-th site in orbital $\mu$ and spin $\sigma$, and $\mu_0$ is the chemical potential used to set the doping $\delta=\langle n \rangle - 6.0$.
The indices $\mu$ and $\nu$ run from 1 to 5 corresponding to the Fe orbitals $d_{3z^2-r^2}$, $d_{yz}$, $d_{xz}$, $d_{xy}$, and $d_{x^2-y^2}$, respectively. The presence of an external magnetic field is described by the Peierls phases
$\varphi_{\mathbf{ij}}=\frac{-\pi}{\Phi_{0}}\oint_{\mathbf j}^{\mathbf i} \mathbf{A} \cdot  \mathbf{dr}$, 
where $\Phi_{0}=\frac{h}{2e}$ is the half flux quantum and the integral is a line integral along the straight line joining lattice sites
$\mathbf{j}$ and $\mathbf{i}$.

The second term in Eq.~(\ref{eq:H}) describes the on-site Coulomb interaction
\begin{align}
 \label{eq:Hint}
 H_{int}&=U\sum_{\mathbf{i},\mu}n_{\mathbf{i}\mu\uparrow}n_{\mathbf{i}\mu\downarrow}+(U'-\frac{J}{2})\sum_{\mathbf{i},\mu<\nu,\sigma\sigma'}n_{\mathbf{i}\mu\sigma}n_{\mathbf{i}\nu\sigma'}\\\nonumber
&\quad-2J\sum_{\mathbf{i},\mu<\nu}\mathbf{S}_{\mathbf{i}\mu}\cdot\mathbf{S}_{\mathbf{i}\nu}+J'\sum_{\mathbf{i},\mu<\nu,\sigma}c_{\mathbf{i}\mu\sigma}^{\dagger}c_{\mathbf{i}\mu\bar{\sigma}}^{\dagger}c_{\mathbf{i}\nu\bar{\sigma}}c_{\mathbf{i}\nu\sigma},
\end{align}

which includes the intraorbital (interorbital) interaction $U$ ($U'$), the Hund's rule coupling $J$ and the pair hopping energy $J'$.
We assume orbitally rotation-invariant interactions $U'=U-2J$ and $J'=J$.

The third term in Eq.~(\ref{eq:H}) contains the superconducting pairing 
\begin{equation}
 H_{BCS}=-\sum_{\mathbf{i}\neq \mathbf{j},\mu\nu}[\Delta_{\mathbf{ij}}^{\mu\nu}c_{\mathbf{i}\mu\uparrow}^{\dagger}c_{\mathbf{j}\nu\downarrow}^{\dagger}+\mbox{H.c.}],
\end{equation}
with gap function $\Delta_{\mathbf{ij}}^{\mu\nu}=\sum_{\alpha\beta}\Gamma_{\mu\alpha}^{\beta\nu}(\mathbf{r_{ij}})\langle\hat{c}_{\mathbf{j}\beta\downarrow}\hat{c}_{\mathbf{i}\alpha\uparrow}\rangle$.
Here $\Gamma_{\mu\alpha}^{\beta\nu}(\mathbf{r_{ij}})$ denotes the effective pairing strength between sites (orbitals) $\mathbf{i}$ and $\mathbf{j}$ ($\mu$, $\nu$, $\alpha$ and $\beta$) connected by vector $\mathbf{r_{ij}}$.
We include only the $\Gamma_{\mu \mu}^{\mu \mu}$ and $\Gamma_{\mu \nu}^{\nu \mu}$ terms coupling next-nearest neighbors (which gives rise to an $s_{\pm}$ pairing state) with the same magnitude $\Gamma_{\mu \mu}^{\mu \mu}=\Gamma_{\mu \nu}^{\nu \mu}=0.117$~eV. This particular choice of pairing terms leads to substantial competition of superconductivity and magnetism in the coexistence phase. A sufficiently strong competition in agreement with experiments is {\it not} found when using the standard couplings generated from the normal state RPA spin fluctuations. This is a strong indication that the pairings below T$_N$ is strongly affected by the reconstructed Fermi surface similar to what is known for the cuprates.\cite{schrieffer89} 

The last term in Eq. (\ref{eq:H}), $H_Z$, is the Zeeman term accounting for energy shifts due to the external magnetic field
\begin{equation}
H_{Z}= h \sum_{\mu} (n_{i \mu \uparrow}-n_{i \mu \downarrow}).
\end{equation}
Here, $ h=-\frac{\mu_{B}g_{s}B}{2}$, $\mu_B$ is the Bohr magneton and $g_s$ is electron g-factor.

A mean-field decoupling of Eq.~\eqref{eq:Hint} leads to the following multi-band Bogoliubov de-Gennes (BdG) equations\cite{gastiasoro13,gastiasoro_jsnm}

\begin{align}
\sum_{\mathbf{j}\nu}
\begin{pmatrix}
H^{\mu\nu}_{\mathbf{i} \mathbf{j} \sigma} & \Delta^{\mu\nu}_{\mathbf{i} \mathbf{j}}\\
\Delta^{\mu\nu*}_{\mathbf{i} \mathbf{j}} & -H^{\mu\nu*}_{\mathbf{i} \mathbf{j} \bar{\sigma}}
\end{pmatrix}
\begin{pmatrix}
 u_{\mathbf{j}\nu}^{n} \\ v_{\mathbf{j}\nu}^{n}
\end{pmatrix}=E_{n}
\begin{pmatrix}
 u_{\mathbf{i}\mu}^{n} \\ v_{\mathbf{i}\mu}^{n}
\end{pmatrix},
\end{align}
where
\begin{align}
 H^{\mu\nu}_{\mathbf{i} \mathbf{j} \sigma}&=t_{\mathbf{ij}}^{\mu\nu} e^{i \varphi_\mathbf{ij}}+\delta_{\mathbf{ij}}\delta_{\mu\nu}[-\mu_0+U \langle n_{\mathbf{i}\mu\bar{\sigma}}\rangle\\\nonumber
\quad&+\sum_{\mu' \neq \mu}(U'\langle n_{\mathbf{i}\mu' \bar{\sigma}}\rangle+(U'-J)\langle n_{\mathbf{i}\mu' \sigma}\rangle)+h].
 \end{align}
We find the stable solutions through iterations of the following self-consistency equations 
\begin{eqnarray}
\langle n_{\mathbf{i}\mu\uparrow} \rangle&=&\sum_{n}|u_{\mathbf{i}\mu}^{n}|^{2}f(E_{n}),\\ 
\langle n_{\mathbf{i}\mu\downarrow} \rangle\!&=&\!\sum_{n}|v_{\mathbf{i}\mu}^{n}|^{2}(1\!-\!f(E_{n})),\\
\Delta_{\mathbf{ij}}^{\mu\nu}&=&\sum_{\alpha\beta}\Gamma_{\mu\alpha}^{\beta\nu}(\mathbf{r_{ij}})\sum_{n}u_{\mathbf{i}\alpha}^{n}v_{\mathbf{j}\beta}^{n*}f(E_{n}), 
\end{eqnarray}
where $\sum_n$ denotes summation over all eigenstates $n$.

Below, the lattice constant $a$ is chosen as the unit of length, and we apply the Landau gauge $\mathbf{A}(\mathbf{r})=(By,0)$, which corresponds to a magnetic field $\mathbf{B}=B (-\mathbf{\hat e_{z}})$. We introduce the magnetic translation operators (MTO) of the Bogoliubov quasiparticles, which commute with the Hamiltonian, as follows
\begin{equation}
\label{MTO}
 \mathcal{M}_{\mathbf R} \hat \psi(\mathbf r)=e^{-i \frac{1}{2} \chi(\mathbf r,\mathbf R ) \sigma_{z}} \hat \psi(\mathbf r - \mathbf R ),
\end{equation}
where $\sigma_{z}$ is the Pauli matrix, $\hat \psi(\mathbf r)$ are the wave functions of the quasiparticles with $u$ and $v$ components, $\chi(\mathbf r,\mathbf R)=\frac{2 \pi}{\Phi_{0}} \mathbf A (\mathbf R) \cdot \mathbf r$ and $\mathbf R= m N_{x} \mathbf{\hat e_{x}}+n N_{y} \mathbf{\hat e_{y}}$ with $m,n$ integers and $N_{x}$,$N_{y}$ being the dimensions of the magnetic unit cell (MUC). 
In order to have MTOs that fulfill the composition law $\mathcal{M}_{\mathbf R_{m}} \mathcal{M}_{\mathbf R_{n}}= \mathcal{M}_{\mathbf R_{m} + \mathbf R_{n}}$, it is required that the MUC is crossed by an even number of half flux quantum $\Phi_{0}$. The magnetic field is fixed such that the flux going through the MUC is $\Phi=2 \Phi_{0}$. The fulfillment of the composition law leads to the generalized Bloch theorem, which reads
\begin{equation}
\label{Bloch}
 \mathcal{M}_{\mathbf R} \hat{\psi}_{\mathbf{k}}(\mathbf r)= e^{-i \mathbf k \cdot \mathbf R} \hat{\psi}_{\mathbf{k}}(\mathbf r),
\end{equation}
where $\mathbf k $=$\frac{2 \pi l_{x}}{N_{x}} \mathbf{\hat e_{x}}+\frac{2 \pi l_{y}}{N_{y}} \mathbf{\hat e_{y}}$ with $l_{x,y}=0,1,...,N_{x,y}-1$ are the wave vectors defined in the first Brillouin zone of the vortex lattice and $ \hat{\psi}_{k}(\mathbf r)$ denote eigenstates of the Hamiltonian and the MTO. By use of Eq.~(\ref{MTO}) and Eq.~(\ref{Bloch}), the eigenfunctions of the Hamiltonian transform under translations as
\begin{align}
\begin{pmatrix}
 u_{\mathbf{i+R}\mu}^{n} \\ v_{\mathbf{i+R}\mu}^{n}
\end{pmatrix}
=e^{i \mathbf k \cdot \mathbf R}
\begin{pmatrix}
e^{-i \frac{1}{2}\chi(\mathbf r,\mathbf R )} u_{\mathbf{i}\mu}^{n} \\ e^{i \frac{1}{2} \chi(\mathbf r,\mathbf R )} v_{\mathbf{i}\mu}^{n}
\end{pmatrix},
\end{align}
where $\mathbf i$ takes values in the magnetic unit cell and $\mathbf r$=$\mathbf{i+R}$.
 Note that since a minimum of two superconducting flux quanta need to penetrate the MUC, the magnetic field is related to the real-space system size by $B \sim \frac{46000}{N_{x}N_{y}}T$ (for $a=3 \text{\AA}$). For the five-band model used here, we are restricted numerically to systems of sizes less than $(N_{x},N_{y})=(46,23)$, indicating that we have minimum field of $43.5$ T or larger, which is more than 3 times the field used in the experiments. This, however, is only a quantitative issue and does not influence the main points of the theoretical results discussed in the following sections.

\subsection{Results}

\begin{figure}[t]
\includegraphics[width=8.5cm]{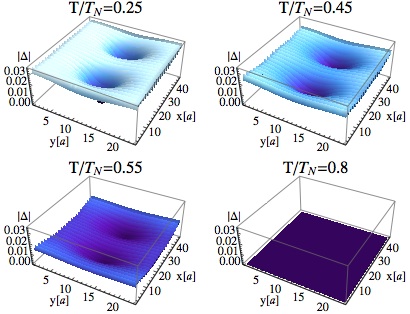}
\caption{Real-space plot of the amplitude of the superconducting order parameter $|\Delta(\mathbf{r_i})|$ for different temperatures $T/T_N$ and constant field $B=43.5$ T. The cores exhibit the usual suppression of 
$|\Delta(\mathbf{r_i})|$.}
\label{fig:4}
\end{figure}

\begin{figure}[b]
\includegraphics[width=9.0cm]{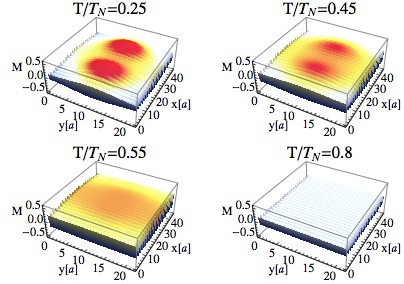}
\caption{Real-space plot of the total magnetization $M(\mathbf{r_{i}})[\mu_B]$ for different temperatures $T/T_N$ and constant field $B=43.5$ T. The vortices are clearly seen to nucleate magnetic order. }
\label{fig:2}
\end{figure}

\begin{figure}[]
\includegraphics[width=8.5cm]{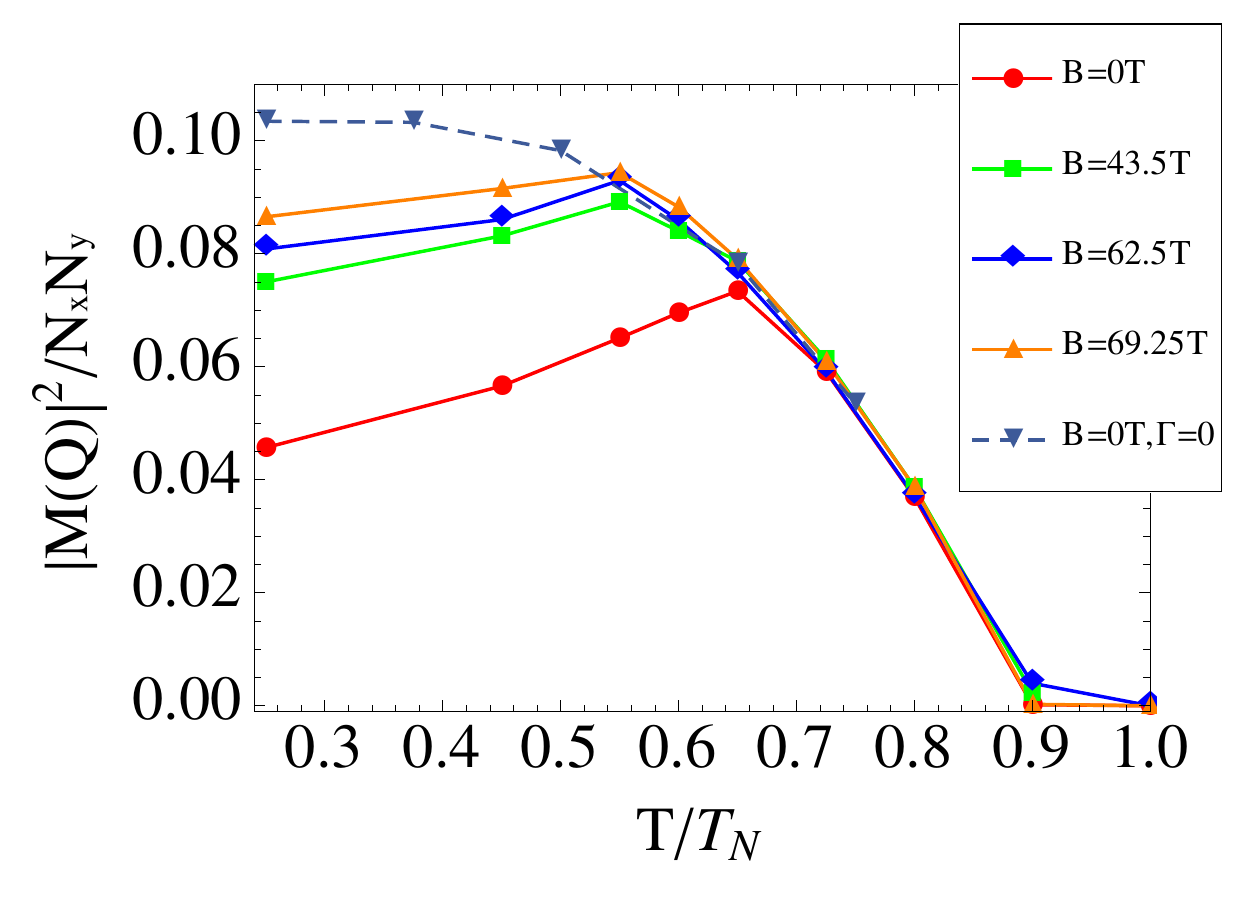}
\caption{Amplitude of the $(\pi,0)$ peak of the Fourier transform of the magnetization $M(\mathbf r)[\mu_{B}]$ versus temperature $T/T_{N}$ for different values of the magnetic field $B$.}
\label{fig:1}
\end{figure}

Figure~\ref{fig:4} shows the self-consistent superconducting order parameter and its temperature dependence for fixed value of the external magnetic field $B=43.5T$. At low $T$, the vortices generated by the external field can be clearly seen. In Fig.~\ref{fig:4} the superconducting order parameter at each site is defined as

\begin{equation}
\Delta(\mathbf{r_i})= \frac{1}{9}\sum_{\mu \nu,\mathbf{j*}} \Delta_{\mathbf{ij*}}^{\mu\nu}e^{i \varphi_{\mathbf{ij*}}},
\end{equation}
where the index $\mathbf{j*}$ includes the set of onsite, nearest neighbor and next-nearest neighbor sites to site $\mathbf{i}$.
With increasing $T$, the amplitude of the superconducting order parameter $|\Delta|$ naturally decreases
and also there is an extra modulation due to an induced stripe structure of the magnetization.

Figure~\ref{fig:2} shows explicitly how the total real space magnetization $M(\mathbf{r_{i}})=\sum_{\mu}(\langle n_{\mathbf{i}\mu\uparrow} \rangle - \langle n_{\mathbf{i}\mu\downarrow} \rangle)$,
changes with temperature for parameters corresponding to Fig.~\ref{fig:4}. As seen, at low temperatures where the superconductivity is strongest, the magnetization is enhanced in the vortices, and as the temperature increases and the amplitude of the superconducting order is reduced, the magnetization enhancement is equally suppressed.  For the parameters used here, the magnetization is enhanced in the region around the vortices below $T/T_{N} \sim 0.56$, which is slightly below the superconducting critical transition temperature $T_c/T_N \sim 0.60$. Above $T_c$, the magnetization decreases homogeneously as the temperature is increased, in agreement with the experimental findings of Fig.~\ref{CompetitionOrder}. The existence of vortex-pinned magnetic order has been studied extensively in the case of cuprates.\cite{andersen00,Chen02,Zhu,Takigawa,Schmid10,Andersen11}  Typically, the suppression of the superconducting order parameter generates low energy vortex core bound states inside the superconducting gap, and the non-interacting spin susceptibility is enhanced at low energies. In conjunction with Coulomb correlations, the latter may lead to local Stoner instabilities making it energetically favorable for the system to nucleate magnetic order locally.\cite{andersen00,Chen02,Zhu,Takigawa,Schmid10,Andersen11} A similar local pinning to magnetic fluctuations is known to take place near potential impurity scatterers in correlated systems.\cite{Tsuchiura,Wang,zhu02,Chen,Kontani,Harter,Andersen07,Andersen10,Christensen} We note that the length scale of the magnetic order pinned by vortices is not given by the superconducting coherence length, but rather determined by the electron interactions and the original band structure.

For direct comparison to the experimental results, we show in Fig.~\ref{fig:1} the square of the amplitude of the Fourier component $|M(\mathbf{Q})|^{2}$ of the magnetization at $\mathbf{Q}=(\pi,0)$,  where $M(\mathbf{q})=1/\sqrt{N_{x}N_{y}} \sum_{i} M(\mathbf{r_i}) e^{-i \mathbf{q}\cdot \mathbf{r_i}}$. This gives a measure of the $(\pi,0)$ stripe order versus temperature $T$. As seen, for the case of zero field $B=0$ T, when superconductivity sets in at $T_c$, the magnetization gets suppressed and reaches at low $T$ approximately half  its value compared to the case without superconductivity, $\Gamma=0$~eV, shown by the dashed line in Fig.~\ref{fig:1}. With increasing $T$ (but still below $T_c$), the amplitude of the superconducting order parameter decreases, as seen from Fig.~\ref{fig:4}. Thus, the suppression exerted by superconductivity on the magnetization decreases due to their competitive interplay, and leads to an increase of the moments until $T_{c}$ is reached as seen from Fig.~\ref{fig:2}. Above $T_
c$, the magnetization decreases in a conventional mean-field fashion for a second order phase transition. Thus, in this way magnetization and superconductivity both coexist and compete below $T_{c}$, similar to the behavior found in experiments as seen in Fig.~\ref{CompetitionOrder}.
Fig.~\ref{fig:1} additionally clearly exhibit how the $(\pi,0)$ magnetic order is enhanced by a magnetic field, with a concomitant reduction of $T_c$.

\section{Conclusions}

In this paper we have reported a field dependence of the previously observed, superconductivity suppressed, but field enhanced, magnetic signal below $T_{c}$, as seen from Fig. \ref{CompetitionOrder}. 

Our neutron scattering data clearly demonstrates a field-enhanced SDW state, competing with and suppressed by superconductivity, in the Co-doped Ba-122 superconductor with $x = 0.05$. This material is known to display both magnetic and superconducting properties consistent with bulk phenomena in both cases. This suggests that the cause of the competition and the field-enhancement should be found on the sub-nanometer scale. We observe no changes of the SDW peak shape which could otherwise explain the field enhancement.

Our transverse field $\mu$SR measurement shows that essentially the whole sample is magnetic and at the same time we also observe signs of a bulk superconductor with a strongly pinned vortex lattice, as seen from Fig. \ref{MuonData}. We observe no change of the magnetic volume fraction below $T_c$ that could explain the suppression of the magnetic Bragg peak intensity observed by neutron diffraction. This suggests that the explanation should be found in a reduction of the magnetic order parameter (the average sublattice magnetic moment).

Finally, we have provided an explicit calculation of the magnetization within a five-band Hubbard model relevant to the iron pnictides. By including superconductivity, we have shown how static magnetic order centered at the vortices is induced by an external field. In the modelling we were restricted to larger fields than those applied in the experiments but, as discussed above, this does not qualitatively affect the results presented in this paper, and the theory describes well the field-enhanced magnetic order seen by the experiments.

\section*{Acknowledgements}
The experimental work was supported by the Danish Research Council FNU through DANSCATT.
B.M.A. acknowledges support from a Lundbeckfond fellowship (grant A9318). B.M.U. acknowledges support from a La Caixa scholarship. The experimental work was performed on the neutron facility SINQ and the muon facility S$\mu$S, both at the Paul Scherrer Institut, Villigen, Switzerland. The work at UniFr has been supported by the Schweizer Nationalfonds (SNF) grants 200020-140225 and 200020-153660.


\begin{thebibliography}{99}

\bibitem{scalapino12} D J. Scalapino, Rev. Mod. Phys. {\bf 84}, 1383 (2012).
%
\bibitem{cupratereview1} M. Vojta, Adv. Phys. {\bf 58}, 699 (2009).
%
\bibitem{cupratereview2} J. M. Tranquada, in Handbook of High-Temperature Superconductivity Theory and Experiment, edited by J. R. Schrieffer, (Springer, New York, 2007).
%
\bibitem{LumsdenJPhys2010} M. D. Lumsden and A. D. Christianson, J. Phys.: Condens Mat. {\bf 22}, 203203 (2010).
%
\bibitem{sanna09} S. Sanna, R. De Renzi, G. Lamura, C. Ferdeghini, A. Palenzona, M. Putti, M. Tropeano, and T. Shiroka, Phys. Rev. B {\bf 80}, 052503 (2009).
%
\bibitem{DrewNatureMat2009} A. J. Drew Ch. Niedermayer, P. J. Baker, F. L. Pratt, S. J. Blundell, T. Lancaster, R. H. Liu, G. Wu, X. H. Chen, I. Watanabe, V. K. Malik, A. Dubroka, M.  R\"{o}ssle, K. W. Kim, C. Baines and C. Bernhard, Nature Mat. {\bf 8}, 310 (2009).
%
\bibitem{aczel08} A. A. Aczel, E. Baggio-Saitovitch, S. L. Bud'ko, P. C. Canfield, J. P. Carlo, G. F. Chen, P. Dai, T. Goko, W. Z. Hu, G. M. Luke, J. L. Luo, N. Ni, D. R. Sanchez-Candela, F. F. Tafti, N. L. Wang, T. J. Williams, W. Yu, and Y. J. Uemura, Phys. Rev. B {\bf 78}, 214503 (2008).
%
\bibitem{fukazawa09} H. Fukazawa, T. Yamazaki, K. Kondo, Y. Kohori, N. Takeshita, P. M. Shirage, K. Kihou, K. Miyazawa, H. Kito, H. Eisaki, and A. Iyo, J. Phys. Soc. Jpn. {\bf 78}, 033704 (2009).
%
\bibitem{park09}J. T. Park, D. S. Inosov, Ch. Niedermayer, G. L. Sun, D. Haug, N. B. Christensen, R. Dinnebier, A. V. Boris, A. J. Drew, L. Schulz, T. Shapoval, U. Wolff, V. Neu, Xiaoping Yang, C. T. Lin, B. Keimer, and V. Hinkov, Phys. Rev. Lett. {\bf 102}, 117006 (2009).
%
\bibitem{goko09} T. Goko, A. A. Aczel, E. Baggio-Saitovitch, S. L. Budko, P. C. Canfield, J. P. Carlo, G. F. Chen, Pengcheng Dai, A. C. Hamann, W. Z. Hu, H. Kageyama, G. M. Luke, J. L. Luo, B. Nachumi, N. Ni, D. Reznik, D. R. Sanchez-Candela, A. T. Savici, K. J. Sikes, N. L. Wang, C. R. Wiebe, T. J. Williams, T. Yamamoto, W. Yu, and Y. J. Uemura, Phys. Rev. B {\bf 80}, 024508 (2009).
%
\bibitem{laplace09} Y. Laplace, J. Bobroff, F. Rullier-Albenque, D. Colson, and A. Forget, Phys. Rev. B {\bf 80}, 140501 (2009).
%
\bibitem{bernhard09} C. Bernhard, A. J. Drew, L. Schulz, V. K. Malik, M. R\"{o}ssle, Ch. Niedermayer, T. Wolf, G. D. Varma, G. Mu, H.-H. Wen, H. Liu, G. Wu, and X. H. Chen, New J. Phys. {\bf 11}, 055050 (2009).
%
\bibitem{julien09} M. H. Julien, H. Mayaffre, M. Horvatic, C. Berthier, X. D. Zhang, W. Wu, G. F. Chen, N. L. Wang, and J. L. Luo, Europhys. Lett. {\bf 87} 37001 (2009).
%
\bibitem{marsik10} P. Marsik, K. W. Kim, A. Dubroka, M. R\"{o}ssle, V. K. Malik, L. Schulz, C. N. Wang, Ch. Niedermayer, A. J. Drew, M. Willis, T. Wolf, and C. Bernhard, Phys. Rev. Lett. {\bf 105}, 057001 (2010).
%
\bibitem{HChen09} H. Chen, Y. Ren, Y. Qiu, W. Bao, R. H. Liu, G. Wu, T. Wu, Y. L. Xie, X. F. Wang, Q. Huang and X. H. Chen, Europhys. Lett. {\bf 85}, 17006 (2009).
%
\bibitem{wiesenmayer11} E. Wiesenmayer, H. Luetkens, G. Pascua, R. Khasanov, A. Amato, H. Potts, B. Banusch, H. H. Klauss, and D. Johrendt, Phys. Rev. Lett. {\bf 107}, 237001 (2011).
%
\bibitem{li12} Z. Li, R. Zhou, Y. Liu, D. L. Sun, J. Yang, C. T. Lin, and G.-q. Zheng, Phys. Rev. B {\bf 86}, 180501(R) (2012).

\bibitem{pratt09} D- K. Pratt, W. Tian, A. Kreyssig, J. L. Zarestky, S. Nandi, N. Ni, S. L. Bud'ko, P. C. Canfield, A. I. Goldman, and R. J. McQueeney, Phys. Rev. Lett. {\bf 103} 087001 (2009).
%
\bibitem{christianson09} A. D. Christianson, M. D. Lumsden, S. E. Nagler, G. J. MacDougall, M. A. McGuire, A. S. Sefat, R. Jin, B. C. Sales, and D. Mandrus, Phys. Rev. Lett. {\bf 103} 087002 (2009).
%
\bibitem{bernhard12} C. Bernhard, C. N. Wang, L. Nuccio, L. Schulz, O. Zaharko, J. Larsen, C. Aristizabal, M. Willis,
A. J. Drew, G. D. Varma, T. Wolf, and Ch. Niedermayer, Phys. Rev. B {\bf 86}, 184509 (2012).
%
\bibitem{dioguardi13} A. P. Dioguardi, J. Crocker, A. C. Shockley, C. H. Lin, K. R. Shirer, D. M. Nisson, M. M. Lawson, N. apRoberts-Warren, P. C. Canfield, S. L. Budko, S. Ran, and N. J. Curro, Phys. Rev. Lett. {\bf 111} 207201 (2013).
%
\bibitem{Lu14} X. Lu, D. W. Tam, C. Zhang, H. Luo, M. Wang, R. Zhang, L. W. Harriger, T. Keller, B. Keimer, L.-P. Regnault, T. A. Maier, and P. Dai, Phys. Rev. B {\bf 90}, 024509 (2014).
%
\bibitem{tucker12} G. S. Tucker, D. K. Pratt, M. G. Kim, S. Ran, A. Thaler, G. E. Granroth, K. Marty, W. Tian, J. L. Zarestky, M. D. Lumsden, S. L. Bud'ko, P. C. Canfield, A. Kreyssig, A. I. Goldman, and R. J. McQueeney, Phys. Rev. B {\bf 86}, 020503(R) (2012).
%
\bibitem{inosov13} D. S. Inosov, G. Friemel, J. T. Park, A. C. Walters, Y. Texier, Y. Laplace, J. Bobroff, V. Hinkov, D. L. Sun, Y. Liu, R. Khasanov, K. Sedlak, Ph. Bourges, Y. Sidis, A. Ivanov, C. T. Lin, T. Keller, and B. Keimer, Phys. Rev. B {\bf 87}, 224425 (2013).
%
\bibitem{MNG14} M. N. Gastiasoro and B. M. Andersen, Phys. Rev. Lett. {\bf 113}, 067002 (2014).
%
\bibitem{hammerath14} F. Hammerath, P. Bonf\'{a}, S. Sanna, G. Prando, R. De Renzi, Y. Kobayashi, M. Sato, and P. Carretta,  Phys. Rev. B {\bf 89}, 134503 (2014).
%
\bibitem{parker09} D. Parker, M. G. Vavilov, A. V. Chubukov, and I. I. Mazin, Phys. Rev. B {\bf 80}, 100508(R) (2009).
%
\bibitem{maiti12} S. Maiti, R. M. Fernandes, and A. V. Chubukov, Phys. Rev. B {\bf 85}, 144527 (2012).
%
\bibitem{vorontsov09} A. B. Vorontsov, M. G. Vavilov, and A. V. Chubukov, Phys. Rev. B {\bf 79}, 060508(R) (2009).
%
\bibitem{vorontsov10} A. B. Vorontsov, M. G. Vavilov, and A. V. Chubukov, Phys. Rev. B {\bf 81}, 174538 (2010).
%
\bibitem{fernandes10} R. M. Fernandes, D. K. Pratt, W. Tian, J. Zarestky, A. Kreyssig, S. Nandi, M.G. Kim, A. Thaler, N. Ni, P.C. Canfield, R. J. McQueeney, J. Schmalian, and A. I. Goldman, Phys. Rev. B {\bf 81}, 140501(R) (2010).
%
\bibitem{nandi10} S. Nandi, M. G. Kim, A. Kreyssig, R. M. Fernandes, D. K. Pratt, A. Thaler, N. Ni, S. L. Budko, P. C. Canfield, J. Schmalian, R. J. McQueeney, and A. I. Goldman, Phys. Rev. Lett. {\bf 104}, 057006 (2010).
%
\bibitem{fernandes13} R. M. Fernandes, S. Maiti, P. W\"{o}lfle, and A. V. Chubukov, Phys. Rev. Lett. {\bf 111}, 057001 (2013).
 %
\bibitem{schrieffer89} J. R. Schrieffer, X. G. Wen, and S. C. Zhang, Phys. Rev. B {\bf 39}, 11 663 (1989).
%
\bibitem{schmidt14} J. Schmiedt, P. M. R. Brydon, and C. Timm, Phys. Rev. B {\bf 89}, 054515 (2014).
%
\bibitem{Lake02} B. Lake, H. M. R\o nnow, N. B. Christensen, G. Aeppli, K. Lefmann, D. F. McMorrow, P. Vorderwisch, P. Smeibidl, N. Mangkorntong, T. Sasagawa, M. Nohara, H. Takagi, and T. E. Mason, Nature (London) {\bf 415}, 299-302 (2002).
%
\bibitem{Chang08} J. Chang, Ch. Niedermayer, R. Gilardi, N. B. Christensen, H. M. R\o nnow, D. F. McMorrow, M. Ay, J. Stahn, O. Sobolev, A. Hiess, S. Pailh\'{e}s, C. Baines, N. Momono, M. Oda, M. Ido, and J. Mesot, Phys. Rev. B {\bf 78}, 104525 (2008).
%
\bibitem{Kofu09} M. Kofu, S.-H. Lee, M. Fujita, H.-J. Kang, H. Eisaki, and K. Yamada, Phys. Rev. Lett. \textbf{102}, 047001 (2009).
%
\bibitem{Chang09} J. Chang, N. B. Christensen, Ch. Niedermayer, K. Lefmann, H. M. R\o nnow, D. F. McMorrow, A. Schneidewind, P. Link, A. Hiess, M. 
Boehm, R. Mottl, S. Pailh\'{e}s, N. Momono, M. Oda, M. Ido, and J. Mesot, Phys. Rev. Lett. {\bf 102}, 177006 (2009).
%
\bibitem{Khaykovich} B. Khaykovich, S. Wakimoto, R.J. Birgeneau, M. A. Kastner, Y. S. Lee, P. Smeibidl, P. Vorderwisch, and K. Yamada, Phys. Rev. B \textbf{71}, R220508 (2005).
%
\bibitem{zhao10} J. Zhao, L.-P. Regnault, C. Zhang, M. Wang, Z. Li, F. Zhou, Z. Zhao, C. Fang, J. Hu, and P. Dai, Phys. Rev. B {\bf 81}, 180505(R) (2010).
%
\bibitem{li11} S. Li, X. Lu, M. Wang, H.-q. Luo, M. Wang, C. Zhang, E. Faulhaber, L.-P. Regnault, D. Singh, and P. Dai, Phys. Rev. B {\bf 84}, 024518 (2011).
%
\bibitem{wang11} M. Wang, H. Luo, M. Wang, S. Chi, J. A. Rodriguez-Rivera, D. Singh, S. Chang, J. W. Lynn, and P. Dai, Phys. Rev. B {\bf 83}, 094516 (2011).
%
\bibitem{wen10} J. S. Wen, G. Y. Xu, Z .J. Xu, Z. W. Lin, Q. Li, Y. Chen, S. X. Chi, G. D. Gu, and J. M. Tranquada, Phys. Rev. B {\bf 81}, 100513 (2010).
%
\bibitem{ChuPRB2009} J. H. Chu, J. G. Analytis, C. Kucharczyk, and I. R. Fisher, Phys. Rev. B {\bf 79}, 014506 (2009).
%
\bibitem{NiPRB2008} N. Ni, M. E. Tillman, J.-Q. Yan, A. Kracher, S. T. Hannahs, S. L. Bud'ko, and P. C. Canfield, Phys. Rev. B 78, 214515 (2008).
%
\bibitem{Lefmann06} K. Lefmann, Ch. Niedermayer, A. B. Abrahamsen, C. R. H. Bahl, N. B. Christensen, H. S. Jacobsen, T. L. Larsen, P. S. H\"afliger, U. Filges, and H. M. R\o nnow, Physica B {\bf 385-386}, 1083 (2006).
%
\bibitem{Bahl06} C. R. H. Bahl, K. Lefmann, A. B. Abrahamsen, H. M. R\o nnow, F. Saxild, T. B. S. Jensen, L. Udby, N. H. Andersen, N. B. Christensen, H. S. Jakobsen, T. Larsen, P. S. H\"afliger, S. Streule, Ch. Niedermayer, Nucl. Instr. Meth. B {\bf 246}, 2, 452 (2006).
%
\bibitem{RotterPRB} M. Rotter, M. Tegel, D. Johrendt, I. Schellenberg, W. Hermes, and R. P\"ottgen, Phys. Rev. B {\bf 78}, 020503(R) (2008).
%
\bibitem{MaeterPRB09} H. Maeter, H. Luetkens, Yu. G. Pashkevich, A. Kwadrin, R. Khasanov, A. Amato, A. A. Gusev, K. V. Lamonova, D. A. Chervinskii, R. Klingeler, C. Hess, G. Behr, B. B\"uchner, and H.-H. Klauss, Phys. Rev. B {\bf 80}, 094524 (2009).
%
\bibitem{ikeda10} H. Ikeda, R. Arita, and J. Kunes, Phys. Rev. B {\bf 81}, 054502 (2010).
%
\bibitem{gastiasoro13} M. N. Gastiasoro, P. J. Hirschfeld, and B. M. Andersen, Phys. Rev. B {\bf 88}, 220509(R) (2013).
%
\bibitem{gastiasoro_jsnm} M. N. Gastiasoro and B. M. Andersen, J. Supercond. Novel. Magn. {\bf 26}, 2561 (2013).
%
\bibitem{andersen00} B. M. Andersen, H. Bruus, and P. Hedeg\aa rd, Phys. Rev. B \textbf{61}, 6298 (2000).
%
\bibitem{Chen02} Y. Chen and C. S. Ting, Phys. Rev. B \textbf{65}, 180513(R) (2002).
%
\bibitem{Zhu} J. X. Zhu, I. Martin, and A. R. Bishop, Phys. Rev. Lett. \textbf{89}, 067003 (2002).
%
\bibitem{Takigawa} M. Takigawa, M. Ichioka, and K. Machida, Phys. Rev. Lett. \textbf{90}, 047001 (2003).
%
\bibitem{Schmid10} M. Schmid, B. M. Andersen, A. P. Kampf, and P. J. Hirschfeld, New J. Phys. \textbf{12}, 053043 (2010).
%
\bibitem{Andersen11} B. M. Andersen, S. Graser, M. Schmid, A. P. Kampf, and P. J. Hirschfeld, J. Phys. Chem. Solids \textbf{72}, 358 (2011).
%
\bibitem{Tsuchiura} H. Tsuchiura, Y. Tanaka, M. Ogata, and S. Kashiwaya, Phys. Rev. B {\bf 64}, 140501(R) (2001).
%
\bibitem{Wang} Z. Wang and P. A. Lee, Phys. Rev. Lett. 89, 217002 (2002). 
%
\bibitem{zhu02} J.-X. Zhu, I. Martin, and A. R. Bishop, Phys. Rev. Lett. {\bf 89}, 067003 (2002).
%
\bibitem{Chen} Y. Chen and C. S. Ting, Phys. Rev. Lett. {\bf 92}, 077203 (2004).
%
\bibitem{Kontani} H. Kontani and M. Ohno, Phys. Rev. B {\bf 74}, 014406 (2006).
%
\bibitem{Harter} J. W. Harter, B. M. Andersen, J. Bobroff, M. Gabay, and P. J. Hirschfeld, Phys. Rev. B {\bf 75}, 054520 (2007).
%
\bibitem{Andersen07} B. M. Andersen, P. J. Hirschfeld, A. P. Kampf, and M. Schmid, Phys. Rev. Lett. {\bf 99}, 147002 (2007).
%
\bibitem{Andersen10} B. M. Andersen, S. Graser, and P. J. Hirschfeld, Phys. Rev. Lett. {\bf 105}, 147002 (2010).
%
\bibitem{Christensen} R. B. Christensen, P. J. Hirschfeld, and B. M. Andersen, Phys. Rev. B {\bf 84}, 184511 (2011).
%
\bibitem{HuangPRL} Q. Huang, Y. Qiu, W. Bao, M. A. Green, J. W. Lynn, Y. C. Gasparovic, T. Wu, G. Wu, and X. H. Chen, Phys. Rev. Lett {\bf 101}, 257003 (2008).




 
\end{thebibliography}
\end{document}